# Theory of valley-dependent transport in graphene-based lateral quantum structures

Feng-Wu Chen[1], Mei-Yin Chou[2,3,4], Yiing-Rei Chen[5], and Yu-Shu Wu[1,6,*]

[1] Physics Department, National Tsing-Hua University, Hsin-Chu 30013, Taiwan
[2]Institute of Atomic and Molecular Science, Academia Sinica, Taipei 10617, Taiwan
[3]Department of Physics, National Taiwan University, Taipei 10617, Taiwan
[4]School of Physics, Georgia Institute of Technology, Atlanta, GA 30330, USA
[5]Physics Department, National Taiwan Normal University, Taipei 11650, Taiwan
[6]Department of Electrical Engineering, National Tsing-Hua University, Hsin-Chu 30013, Taiwan
[*]Corresponding author. Email: yswu@ee.nthu.edu.tw

**Abstract**

Modulation of electronic states in two-dimensional (2D) materials can be achieved by using in-plane variations of the band gap or the average potential in lateral quantum structures. In the atomic configurations with hexagonal symmetry, this approach makes it possible to tailor the valleytronic properties for potential device applications. In this work, we present a multi-band theory to calculate the valley-dependent electron transport in graphene-based lateral quantum structures. As an example, we consider the structures with a single interface that exhibits an energy gap or potential discontinuity. The theoretical formalism proceeds within the tight-binding description, by first deriving the local bulk complex band structures in the regions of a constant gap or potential and, next, joining the local wave functions across the interface via a cell-averaged current operator to ensure the current continuity. The theory is applied to the study of electron reflection off and transmission through an interface. Both reflection and transmission are found to exhibit valley-contrast behavior that can be used to generate valley-polarized electron sources. The results vary with the type of interfaces, as well as between monolayer and bilayer graphene based structures. In the monolayer case, the valley contrast originates from the band warping and only becomes sizable for incident carriers of high energy; whereas in AB-stacked bilayer graphene, the vertical interlayer coupling emerges as an additional important cause for valley contrast, and the favorable carrier energy is also found to be drastically lower. Our numerical results clearly demonstrate the propitious valleytronic properties of bilayer graphene structures.

## I. Introduction

Electrons in two-dimensional (2D) hexagonal materials such as graphene[1-4] and TMDs[5-8] carry a novel degree of freedom - valley pseudospin, in association with the two-fold valley degeneracy existing in the band structure at points K and K′ of the Brillouin zone. Being binary valued, a valley pseudospin can play the role of an information carrier besides spin and charge, with a favorable advantage that the large wave vector difference between K and K′ effectively suppresses the intervalley scattering and preserves the valley coherence. This makes viable the implementation of valley pseudospin-based devices known as valleytronics for future electronics. Examples of valleytronic devices that have been proposed include valley filters[1,9-11], qubits[12,13], and FETs[13,14].

In valleytronics, it would be of great value to be able to tailor valley-dependent electronic properties in materials to suit applications. This control can be achieved with, for example, the utilization of graphene-based lateral quantum structures with in-plane potential variations ($\delta V$) or gap variations ($\delta \Delta$) to modulate the electronic bands. According to previous studies[12-14], $\delta V$ ($\delta \Delta$) enables the tuning of valleytronic properties in graphene by interacting with the valley pseudospin via the so-called valley-orbit interaction (VOI) – a valley-dependent term in the electron Hamiltonian. In the case of gapped monolayer graphene, for example, the interaction is given by $H_{VOI} \propto \tau \hat{z} \cdot \vec{p} \times \nabla V$ (or $H_{VOI} \propto \tau \hat{z} \cdot \vec{p} \times \nabla \Delta$), where $\tau$ is the valley index (-1/+1 for K/K′). To quantitatively study this type of VOI associated with $\delta V$ or $\delta \Delta$, we develop in this work a multi-band theory of electronic states for graphene-based lateral quantum structures and evaluate the valley-dependent transport of electrons through an interface.

In order to make the experiment feasible, the systems considered in this work are gated structures in monolayer graphene (MLG) on h-BN or AB-stacked bilayer graphene (BLG), where regions with a constant gap or a constant potential can be realized by applying local vertical gate biases to these two-dimensional materials[15-16], with a bias difference between neighboring zones creating a discontinuity $\delta V$ (or $\delta \Delta$). Following the classification scheme in the field of semiconductor heterostructures, we consider three types of discontinuity interfaces between zones, namely, those of type I with straddling gap alignment, type II with staggered or broken gap alignment, and type III with inverted gap alignment, which are all experimentally accessible in graphene structures with the aforementioned vertical gate biases.

The theoretical presentation is summarized below. We work within the tight-binding (TB) model of graphene and start with an analytical, symmetry-based discussion of valley-dependent electron transport in graphene-based structures. It is then followed by a discussion of the theoretical formalism of a multi-band theory that enables the inclusion of intervalley transfer caused by interface scattering in the theory. Specifically, at the core of the formalism are **a)** an algorithm to calculate the bulk complex band structure and **b)** a current density operator to join local wave functions. The algorithm yields, in regions of a constant gap (potential), bulk electronic states with complex wave vectors in general, which can be linearly combined together to form the local wave functions on one side of the interface. The current operator is applied to enforce, on a cell-averaged basis, the condition of current continuity at the interface when joining the local wave functions in two neighboring zones. Last, we apply the formalism to numerically study valley-dependent transport in structures with one single interface, and investigate the valley contrast in electron reflection off and transmission through the interface. The study has important implications for applications such as the generation of valley polarized electron sources for valleytronic signal processing.

Important findings from our study are summarized below. In the electron reflection off or transmission through the interface, the valley contrast can be created with all three types of interfaces, with the magnitude of contrast dependent on the incident angle. However, significant differences in the contrast are found to exist among different types of interfaces as well as between MLG- and BLG- based structures. For example, with MLG, it shows that the contrast derives uniquely from the energy band warping and thus only becomes sizable for electrons of high incident energy, e.g., a few hundred meV off the band edge, where the warping is significant. A similar conclusion has been reported for warping-based valley filters[10]. However, with AB stacked BLG, it is found that the occurrence of sizable contrast comes down to a much lower energy of the order of 10 meV, making BLG a favorable system for valleytronics. In addition, our analysis indicates that the existence of interlayer coupling in BLG largely alters the underlying mechanism of valley contrast. For example, the removal of band warping terms from the Hamiltonian can still result in a sizable contrast in tunneling transport as opposed to the case of MLG, where the removal would totally eliminate it.

This article is organized as follows. **Section II** presents the multi-band theory. **Section III** discusses the numerical result of electron reflection and transmission.

**Section IV** concludes the study. **Appendices I** and **II** develop, respectively, the form and the orthogonality property of current density operator within the TB model of graphene.

## II. The multi-band tight-binding theory

Both MLG- and BLG- based structures with a single interface of gap/potential discontinuity will be considered. We refer to **Figure 1** for the graphene crystal orientation, lattice vectors, and Brillouin zone. The $C_{A1} - C_{B1}$ bond length is denoted by $a$. $\mathbf{a_1} = \sqrt{3}a(0,1)$ and $\mathbf{a_2} = \sqrt{3}a\left(-\frac{\sqrt{3}}{2}, \frac{1}{2}\right)$ are the primitive lattice vectors. $\mathbf{b_1} = \frac{4\pi}{3a}\left(\frac{1}{2}, \frac{\sqrt{3}}{2}\right)$ and $\mathbf{b_2} = \frac{4\pi}{3a}(-1,0)$ are the primitive reciprocal lattice vectors.

For the presentation, we specifically take the interface to be along the armchair direction and given by, e.g., y = 0. Thus, y < 0 and y > 0 specify the two separate zones of constant energy gap (labeled by $\zeta$), with each being subject to a distinct effective vertical gate voltage $V^\zeta$. For the discussion of electron transport throughout this work, we will always assume **a)** the incident electron comes from $y = -\infty$, **b)** the interface at y = 0 is rectilinear and preserves the x-translational symmetry, and **c)** the effect of inelastic electron scattering is negligible. Excluding the higher-order effects in association with the violation of **b)** and **c)**, we take $k_x$ (wave vector in the x-direction) and $E$ (electron energy) to be the appropriate parameters specifying an interface-scattered electron.

**Subsections II-A** and **II-B** provide analytic symmetry-based discussions of the valley-dependent electron transport for the MLG- and BLG- based structures, respectively, within the linearized model for electrons near the Dirac points $K' = \frac{2}{3}\mathbf{b_1} + \frac{1}{3}\mathbf{b_2}$ and $K = -\frac{2}{3}\mathbf{b_1} - \frac{1}{3}\mathbf{b_2}$. **Subsection II-C** goes beyond the linearized approximation and presents the theoretical formalism of the full TB model, which accounts for interface-induced intervalley scattering and is also well suited to the numerical study of electron transport in the structure of interest here.

In passing, we note that the discussion presented in this section can be easily generalized to the case of the zigzag-cut interface.

## II-A.  MLG-based structures with the linearized model

We first consider the case of monolayer graphene. The tight-binding Hamiltonian in the basis of Bloch states constructed out of the 2p$_z$ orbitals at the A and B sites $(\hbar \equiv 1)$ is given by

$$H^{\zeta}(\mathbf{k}) = \begin{pmatrix} \Delta^{\zeta} + V^{\zeta} & -t\left[1 + 2e^{i3k_x a/2} \cos(\sqrt{3}k_y a/2)\right] \\ -t\left[1 + 2e^{-i3k_x a/2} \cos(\sqrt{3}k_y a/2)\right] & -\Delta^{\zeta} + V^{\zeta} \end{pmatrix} \quad (1)$$

with $-t$ $(t > 0)$ being the nearest neighbor $(p_z p_z \pi)$ hopping energy and $\mathbf{k} = (k_x, k_y)$ the electron wave vector. For each zone $\zeta$, $2\Delta^{\zeta}$ is the possible on-site energy difference between the A and B sites, on top of the average level $V^{\zeta}$. In particular, we take $\Delta^{\mathrm{I}} = \Delta^{\mathrm{II}}$, because the energy gap in MLG is basically gate bias independent. For $\mathbf{k}$ near the Dirac points, the linearized Dirac Hamiltonian is given by[15]:

$$H^{\zeta,\tau}(\mathbf{k}) = \begin{pmatrix} \Delta^{\zeta} + V^{\zeta} & v_F(\tau k_y - ik_x) \\ v_F(\tau k_y + ik_x) & -\Delta^{\zeta} + V^{\zeta} \end{pmatrix} = H_{\mathrm{MLG}}^{\tau}(\mathbf{k}) + \Delta^{\zeta}\sigma_z + V^{\zeta}\mathbf{1} \quad (1')$$

$$\text{where } H_{\mathrm{MLG}}^{\tau}(\mathbf{k}) = v_F(-\sigma_y k_x + \tau\sigma_x k_y) \equiv \begin{pmatrix} 0 & \beta^* \\ \beta & 0 \end{pmatrix}. \quad (2)$$

Now, $\mathbf{k} = (k_x, k_y)$ denotes the electron wave vector relative to the Dirac points, $\tau$ is the valley index, and $v_F$ is the Fermi velocity given by $v_F = \dfrac{3ta}{2}$. In the vicinities of the valley minima $(|\beta| \ll \Delta^{\zeta})$, the eigenvalue and eigenvector of the Hamiltonian are well known and given, respectively, by

$$\varepsilon^{\zeta,\tau}(\mathbf{k}) = V^{\zeta} \pm \left[(\Delta^{\zeta})^2 + |\beta|^2\right]^{\frac{1}{2}} \cong V^{\zeta} \pm \left(\Delta^{\zeta} + \frac{|\beta|^2}{2\Delta^{\zeta}}\right), \quad (3)$$

$$\xi^{\zeta,\tau}(\boldsymbol{k}) = \frac{1}{\Omega^{\zeta,\tau}(\boldsymbol{k})} \begin{pmatrix} \Delta^{\zeta} + \varepsilon^{\zeta,\tau}(\boldsymbol{k}) \\ \beta(\boldsymbol{k}) \end{pmatrix} \equiv \begin{pmatrix} v_A^{\zeta,\tau}(\boldsymbol{k}) \\ v_B^{\zeta,\tau}(\boldsymbol{k}) \end{pmatrix}. \tag{4}$$

Here, $\Omega^{\zeta,\tau}(\boldsymbol{k})$ is the normalization factor. Thus, within the linearized approximation, the local band structure is characterized by cylindrical symmetry (or absence of warping), a shift by $V^{\zeta}$, and a quadratic-in-$|\beta|$ dependence near $|\beta| = 0$ with a gap of $2\Delta^{\zeta}$. For $\boldsymbol{k}' = (k_x, -k_y)$, the eigenvalue remains the same, but the eigenvector becomes

$$\xi^{\zeta,\tau}(\boldsymbol{k}') = \begin{pmatrix} \left(v_A^{\zeta,\tau}\right)^*(\boldsymbol{k}) \\ -\left(v_B^{\zeta,\tau}\right)^*(\boldsymbol{k}) \end{pmatrix}. \tag{5}$$

Moreover, for each $\boldsymbol{k}$, due to the reflection symmetry between y and -y, the following simple relation holds between the corresponding eigenvectors from both valleys, given by

$$\xi^{\zeta,-\tau}(\boldsymbol{k}) = \begin{pmatrix} v_A^{\zeta,-\tau}(\boldsymbol{k}) \\ v_B^{\zeta,-\tau}(\boldsymbol{k}) \end{pmatrix} = \begin{pmatrix} \left(v_A^{\zeta,\tau}\right)^*(\boldsymbol{k}) \\ -\left(v_B^{\zeta,\tau}\right)^*(\boldsymbol{k}) \end{pmatrix} = \xi^{\zeta,\tau}(\boldsymbol{k}'). \tag{6}$$

Now, consider an incident electron of wave vector $\boldsymbol{k}$ and energy $E$ in zone I (y < 0), which is partially reflected off and partially transmitted through the interface (y = 0) into zone II (y > 0). Within the present linearized model, we can construct the local wave function at lattice site R on each side of the interface without intervalley mixing, as given below:

$$\Psi^{I,\tau}(\boldsymbol{k}, \boldsymbol{R}) = e^{i\boldsymbol{k}\cdot\boldsymbol{R}} \begin{pmatrix} v_A^{I,\tau}(\boldsymbol{k}) \\ v_B^{I,\tau}(\boldsymbol{k}) \end{pmatrix} + r^{\tau} e^{i\boldsymbol{k}'\cdot\boldsymbol{R}} \begin{pmatrix} \left(v_A^{I,\tau}\right)^*(\boldsymbol{k}) \\ -\left(v_B^{I,\tau}\right)^*(\boldsymbol{k}) \end{pmatrix}, \tag{7}$$

$$\Psi^{II,\tau}(\boldsymbol{k}, \boldsymbol{R}) = t^{\tau} e^{i\boldsymbol{q}\cdot\boldsymbol{R}} \begin{pmatrix} v_A^{II,\tau}(\boldsymbol{q}) \\ v_B^{II,\tau}(\boldsymbol{q}) \end{pmatrix}. \tag{8}$$

Here, $\boldsymbol{k} = (k_x, k_y)$, $\boldsymbol{k}' = (k_x, -k_y)$ and $\boldsymbol{q} = (k_x, q_y)$, where the component $q_y$ is to be determined by the relative $V^{\zeta}$ and $\Delta^{\zeta}$ values, such that both $k_x$ and $E$ remain invariant across the interface $r^{\tau}$ and $t^{\tau}$ are reflection and transmission parameters for valley $\tau$, respectively.

The continuity condition of $\Psi^\tau$ at the interface determines the reflection and transmission coefficients:

$$\begin{pmatrix} v_A^{I,\tau}(\boldsymbol{k}) \\ v_B^{I,\tau}(\boldsymbol{k}) \end{pmatrix} + r^\tau \begin{pmatrix} \left(v_A^{I,\tau}\right)^*(\boldsymbol{k}) \\ -\left(v_B^{I,\tau}\right)^*(\boldsymbol{k}) \end{pmatrix} = t^\tau \begin{pmatrix} v_A^{II,\tau}(\boldsymbol{q}) \\ v_B^{II,\tau}(\boldsymbol{q}) \end{pmatrix}. \tag{9}$$

Similarly, using the aforementioned relation (6) between corresponding eigenvectors of the two valleys, the continuity condition of $\Psi^{-\tau}$ at the interface gives

$$\begin{pmatrix} \left(v_A^{I,\tau}\right)^*(\boldsymbol{k}) \\ -\left(v_B^{I,\tau}\right)^*(\boldsymbol{k}) \end{pmatrix} + r^{-\tau} \begin{pmatrix} v_A^{I,\tau}(\boldsymbol{k}) \\ v_B^{I,\tau}(\boldsymbol{k}) \end{pmatrix} = t^{-\tau} \begin{pmatrix} \left(v_A^{II,\tau}\right)^*(\boldsymbol{q}) \\ -\left(v_B^{II,\tau}\right)^*(\boldsymbol{q}) \end{pmatrix}. \tag{10}$$

as well. Comparing the two equations above, we find $r^{-\tau} = \left(r^\tau\right)^*$ and $t^{-\tau} = \left(t^\tau\right)^*$, or $\left|r^{-\tau}\right|^2 = \left|r^\tau\right|^2$ and $\left|t^{-\tau}\right|^2 = \left|t^\tau\right|^2$. In other words, there is no valley contrast in the electron reflection or transmission.

In the above argument, we have taken $q_y$ to be real, meaning that the electron energy lies also above the conduction band edge in zone II. In the case where $q_y$ is imaginary, extension of the argument is needed but the same conclusion can be verified to hold.

A similar argument can be made in the case where the interface runs along the zigzag direction. In summary, for MLG based structures, we find a lack of valley contrast for either a zigzag-cut or an armchair-cut interface, within the linearized approximation.

*Effect of warping*

Beyond the linearized approximation, band warping appears and the degeneracy between $(\tau, \boldsymbol{k})$ and $(-\tau, \boldsymbol{k})$ states is lifted (See **Figure 1(c)**), invalidating the above analysis. An insignificant valley contrast, e.g., $10^{-7}$, occurs for low-energy electrons, as numerically demonstrated in **Section III** with the full TB model.

## II-B. BLG-based structures with the linearized model

We focus on the so-called AB-stacked bilayer structure. We include only the $2p_z$ orbitals on A1, B1, A2, and B2 sites,[16] and write below the linearized Hamiltonian

$$H^{\zeta,\tau}(k) = \begin{pmatrix} -\Delta^\zeta + V^\zeta & v_F(\tau k_y - ik_x) & 0 & v_\perp(\tau k_y + ik_x) \\ v_F(\tau k_y + ik_x) & -\Delta^\zeta + V^\zeta & \gamma_1 & 0 \\ 0 & \gamma_1 & \Delta^\zeta + V^\zeta & v_F(\tau k_y - ik_x) \\ v_\perp(\tau k_y - ik_x) & 0 & v_F(\tau k_y + ik_x) & \Delta^\zeta + V^\zeta \end{pmatrix} \quad (11)$$

$\gamma_1$ = the vertical interlayer $C_{A2} - C_{B1}$ coupling, $-\gamma_3$ = the non-vertical interlayer $C_{B2} - C_{A1}$ coupling ($\gamma_1 > 0$ and $\gamma_3 > 0$), $2\Delta^\zeta$ = the interlayer potential difference, and $v_\perp = \frac{3\gamma_3 a}{2}$. Throughout the work, the tight-binding parameters used are: $t$ = 2.8 eV, $g_1$ = 0.4 eV, and $g_3$ = 0.3 eV.

The Hamiltonian is a bit more complicated than the MLG's, and results in the well-known local band structure which is characterized by a band gap of $2\Delta^\zeta$ between the fundamental conduction and valence bands, and the two distant bands located at $\pm\left[\left(\Delta^\zeta\right)^2 + |\gamma_1|^2\right]^{\frac{1}{2}}$ away from the band gap. Again the diagonal term $V^\zeta$ shifts the local energy bands. As opposed to MLG, however, it can be verified that in the linearized approximation the trigonal band warping still exists in BLG, due to the presence of the non-vertical interlayer coupling $\gamma_3$. When we set $\gamma_3 = 0$ (and therefore $v_\perp = 0$), then the energy dispersion becomes[16]

$$E_\pm^{(\chi+1)/2+1} = \pm\left[\Delta^2 + v_F^2 k^2 + \frac{\gamma_1^2}{2} + \frac{\chi\gamma_1^2}{2}\sqrt{1 + \frac{4v_F^2 k^2}{\gamma_1^2} + \frac{16\Delta^2 v_F^2 k^2}{\gamma_1^4}}\right]^{1/2}, \quad (12)$$

where $k^2 = k_x^2 + k_y^2$, $\chi = -1$ for the first conduction ($E_+^{(1)}$) / valence ($E_-^{(1)}$) bands and $\chi = 1$ for the second conduction ($E_+^{(2)}$) / valence ($E_-^{(2)}$) bands, all showing cylindrical symmetry.

Because the interface disrupts the translational symmetry in the y-direction, another new important element comes into play, namely, the interband mixing of electron states caused by the interface scattering, in a model that gives rise to multi-conduction and valence bands such as the present one. For example, for incident electrons of the first conduction band, the mixing leads to the emergence of the second conduction band-derived states. In particular, for the incident electron energy that lies below the second conduction band, such states are characterized by complex $k_y = \alpha_2 + i\kappa_2$. The emergence of these evanescent states in the electron transport has important implications for valley contrast, as discussed below.

*Effect of the vertical interlayer coupling*

We retain the vertical coupling $\gamma_1$ but remove the warping by turning off $\gamma_3$. It follows that the state of $\left(\tau, \boldsymbol{k} = (k_x, k_y)\right)$ is degenerate with that of $\left(-\tau, \boldsymbol{k}^* = (k_x, k_y^*)\right)$, with the following relation between the two corresponding eigenvectors

$$\xi^{\zeta,\tau}(\boldsymbol{k}) = \begin{pmatrix} v_{A1}^{\zeta,\tau}(\boldsymbol{k}) \\ v_{B1}^{\zeta,\tau}(\boldsymbol{k}) \\ v_{A2}^{\zeta,\tau}(\boldsymbol{k}) \\ v_{B2}^{\zeta,\tau}(\boldsymbol{k}) \end{pmatrix} \Rightarrow \xi^{\zeta,-\tau}(\boldsymbol{k}^*) = \begin{pmatrix} \left(v_{A1}^{\zeta,\tau}\right)^*(\boldsymbol{k}) \\ -\left(v_{B1}^{\zeta,\tau}\right)^*(\boldsymbol{k}) \\ -\left(v_{A2}^{\zeta,\tau}\right)^*(\boldsymbol{k}) \\ \left(v_{B2}^{\zeta,\tau}\right)^*(\boldsymbol{k}) \end{pmatrix} \qquad (13)$$

Consider incident electrons of the first conduction band. Let us take $k_y = i\kappa_2$ for the evanescent states emerging from the interface scattering, where the sign of $\kappa_2$ is given such that the state decays exponentially into the distance, e.g., $\kappa_2 > 0$ ($\kappa_2 < 0$) for y > 0 (y < 0). We skip the analytical existence proof of states with $k_y = i\kappa_2$, and refer to the numerical evidence in **Figure 2** below for the vanishing of $\alpha_2$ (real part in $k_y$) for the majority of evanescent states between the first and second conduction bands. For such states, $\boldsymbol{k}^* = (k_x, -i\kappa_2)$ refers to a state that diverges exponentially into the distance, as opposed to the state of $\boldsymbol{k} = (k_x, i\kappa_2)$, and has to be excluded from the wave function in accordance with the usual requirement of wave function convergence. It follows that the symmetry argument employed earlier in the MLG case for proving the lack of valley contrast fails in the present case, since the argument hinges on being able to pair degenerate states $(\tau, \boldsymbol{k})$ and $(-\tau, \boldsymbol{k}^*)$ in the interface-scattered wave functions for the incident states of opposite valleys. Overall, with the emergence of evanescent states, the argument fails thus leaving room for a finite valley contrast.

We note that if we further turn off $\gamma_1$, the resulting Hamiltonian would describe two separate, isolated pieces of MLG at different biases, both governed by the linearized Dirac Hamiltonian of MLG. This would bring us back to the MLG case with the lack of valley contrast. Therefore, we attribute the valley contrast in BLG above as caused by the vertical interlayer hopping $\gamma_1$. As numerically demonstrated in **Section III**, the accompanying contrast can be significant in tunneling transport.

*Effect of warping*

If we include the warping resulting from $v_\perp \neq 0$, it would lift the degeneracy between the $(\tau, \boldsymbol{k})$ and $(-\tau, \boldsymbol{k}^*)$ states and thus alter the degree of valley contrast. For numerical demonstration, see **Section III**.

## II-C.  Full TB formalism for electron transport

The linearized model employed in **Subsections II-A** and **II-B** for the study of electron transport treats the two valleys independently and, thus, totally ignores the interface-induced intervalley mixing. For applications that are concerned with intervalley mixing-caused valley decoherence, the model can be improved by taking into account the full TB Hamiltonian that allows for the occurrence of intervalley transition upon interface scattering. This section develops the improvement on the cellular scale, in the sense that the atomic-scale details are integrated out. More specifically, the treatment ensures the probability current continuity on a cellular rather than atomic site basis. The discussion will focus on BLG-based structures, since it can be easily applied to the MLG case by simply turning off the interlayer coupling.

Overall, due to the interface-induced intervalley and interband mixing, an incident state characterized by ($k_x$, $k_y$, $E$, $\tau$), where ($k_x$, $k_y$) is a real electron wave vector in zone I, can be scattered into the opposite valley or evanescent states with complex wave vectors in the y-direction. This means for given $k_x$ and $E$, in the study of electron transport, the construction of the local wave function in zone $\zeta$ must include in the linear combination all local bulk states characterized by $((k_y)_n^{\zeta,\tau'}; k_x, E)$, e.g.,

$$\Psi^{\zeta}(r; k_x, E) = \sum_{\tau'=\pm 1} \sum_{n=1}^{4} c_n^{\zeta,\tau'} \phi_n^{\zeta,\tau'}(r; k_x, E). \tag{14}$$

$\phi_n^{\zeta,\tau'}$ is the bulk state in zone $\zeta$ and is given by the linear combination of Bloch basis functions at each sublattice, i.e., $\phi_n^{\zeta,\tau'} = \sum_{\mu=A1,A2,B1,B2} \left(\xi_n^{\zeta,\tau'}\right)_\mu \Phi_\mu(\mathbf{k})$, where

$\Phi_\mu(\mathbf{k}) = \frac{1}{\sqrt{N}} \sum_{\mathbf{R}} e^{ik_x R_x} \exp\left[i(k_y)_n^{\zeta,\tau'} R_y\right] \varphi_a(\mathbf{r} - \mathbf{R} - \boldsymbol{\delta}_\mu)$, $\mathbf{k} = (k_x, (k_y)_n^{\zeta,\tau'})$, N is the total number of unit cells, $\mathbf{R}$ runs over all the lattice points, $\mu$ is the index of basis atoms, $\boldsymbol{\delta}_\mu$ is the position vector of the $\mu$-basis atom relative to the lattice point, and $\varphi_a$ is the atomic orbital at each sublattice point. The linear combination expression for the total wave function $\Psi^{\zeta}(r; k_x, E)$ conserves $k_x$ and $E$, but allows the choice of $\tau' = \tau$ or $-\tau$ (valley flip) as well as the y-component $(k_y)_n^{\zeta,\tau'}$ of the electron wave vector to be complex.

We now proceed to develop an algorithm that finds all local bulk states for given $k_x$ and $E$. As opposed to our earlier notation, we now re-define $\boldsymbol{k} = (k_x, k_y)$ as the electron wave vector with respect to the Brillouin zone center, and thus treat bulk states of the two valleys in a unified fashion. With this approach, the valley index will be suppressed below. For simplicity, we shall also drop other subscripts or superscripts throughout the presentation, whenever it is possible to do so without causing confusion.

For the start, we write the bulk Hamiltonian equation in zone $\zeta$, with the full TB Hamiltonian:

$$H^{\zeta}(k_x, (k_y)_n^{\zeta}) \xi_n^{\zeta} = E \xi_n^{\zeta}. \tag{15}$$

The Hamiltonian is given by

$$H^\zeta(k_x, (k_y)_n^\zeta) = H_0^\zeta(k_x) + \eta_n^\zeta H_1(k_x);  \qquad (16)$$

$$H_0^\zeta(k_x) = \begin{pmatrix} -\Delta_\zeta & -t & 0 & -\gamma_3 e^{-i3k_x a} \\ -t & -\Delta_\zeta & \gamma_1 & 0 \\ 0 & \gamma_1 & \Delta_\zeta & -t \\ -\gamma_3 e^{i3k_x a} & 0 & -t & \Delta_\zeta \end{pmatrix}, \qquad (17a)$$

$$H_1(k_x) = \begin{pmatrix} 0 & -te^{-i3k_x a/2} & 0 & -\gamma_3 e^{-i3k_x a/2} \\ -te^{i3k_x a/2} & 0 & 0 & 0 \\ 0 & 0 & 0 & -te^{-i3k_x a/2} \\ -\gamma_3 e^{i3k_x a/2} & 0 & -te^{i3k_x a/2} & 0 \end{pmatrix} \qquad (17b)$$

with

$$\eta_n^\zeta = \left[\lambda_n^\zeta + \left(\lambda_n^\zeta\right)^{-1}\right], \quad \lambda_n^\zeta = \exp\left[i\sqrt{3}(k_y)_n^\zeta a/2\right]. \qquad (18)$$

From the above expressions, we see that the bulk states with $(k_y)_n^\zeta$ and $-(k_y)_n^\zeta$ are degenerate, since $\eta_n^\zeta$ is invariant under the transformation $(k_y)_n^\zeta \to -(k_y)_n^\zeta$. Moreover, by taking the Hermitian conjugate of the Hamiltonian equation, we further deduce that the states with $\left[(k_y)_n^\zeta\right]^*$ and $-\left[(k_y)_n^\zeta\right]^*$ are degenerate with that of $(k_y)_n^\zeta$ as well.

In order to find all bulk states (with real or complex $k_y$) for given $E$ and $k_x$, we transform the Hamiltonian equation to the following equation:

$$H_1(k_x)^{-1}\left[E - H_0^\zeta(k_x)\right]\xi_n^\zeta = \eta_n^\zeta \xi_n^\zeta, \qquad (19)$$

with $\eta_n^\zeta$ being the eigenvalue. The transformation allows us to first obtain $\eta_n^\zeta$ as the eigenvalue from Eq. (19), and then use Eq. (18) to obtain $(k_y)_n^\zeta$. Since the operator in the eigenvalue equation is non-Hermitian, the corresponding eigenvalue $\eta_n^\zeta$ is generally complex, thus allowing us to find both real and complex $(k_y)_n^\zeta$ through the aforementioned transformation.

**Figure 2** presents an example of the bulk complex band structure obtained with the above algorithm for BLG with $k_x = 0.0035/K/$ and $\Delta = 20$ meV. Generally, for a given $(k_x, E)$ set, there are eight solutions of $k_y's$ in all, two for either valley of each energy bands. For example, above the second conduction band edge ~ 400 meV, each of the two conduction bands contribute four solutions of $k_y's$, all being real. However, from about 20 meV to 400 meV between the two conduction bands, for the majority of $E$'s, four of the solutions associated with the first band are real and the other four associated with the second band are complex. As discussed in **Subsection II-B**, the existence of such complex solutions leads to the $\gamma_1$-induced valley contrast in electron transport. However, not all of the complex solutions are used to construct the local wave function on a given side of the interface. In the zone of y > 0 (y < 0), half of the complex solutions have $k_y$'s with negative (positive) imaginary parts. Such solutions have to be excluded from the linear combination of the local wave function to ensure the proper asymptotic behavior. Last, we also note that the graphs of E vs. $\left|\text{Re}(k_y^{norm})\right|$ in **Figures 2(a)** and **2(b)** show a slight asymmetry under the mirror reflection with respect to the line of $\left|\text{Re}(k_y^{norm})\right|=1$. This asymmetry comes from the existence of trigonal band warping around the Dirac point (**Figure 1(c)**).

Next, we discuss the connection of local wave functions across the interface. This is achieved by applying the continuity condition of probability current at the interface [17]. Specifically, we enforce the condition on a cellular basis, by integrating out the current within each unit cell. As demonstrated in **Appendix I**, the integration leads to the cell-averaged current density operator, $J_y$, with the following matrix element

$$\left\langle \phi_{n*}^{\zeta} \left| J_y(\boldsymbol{r}) \right| \phi_m^{\zeta'} \right\rangle \equiv J_{n*,m}^{\zeta,\zeta'}(\boldsymbol{r})$$

$$= \frac{i}{\hbar} \frac{\sqrt{3}a}{8N} \exp\left\{-i\left[\left(k_y\right)_n^{\zeta} - \left(k_y\right)_m^{\zeta'}\right] y\right\}$$

$$\times \left\{\left[\lambda_n^{\zeta} - \left(\lambda_n^{\zeta}\right)^{-1}\right] + \left[\lambda_m^{\zeta'} - \left(\lambda_m^{\zeta'}\right)^{-1}\right]\right\} \left(\xi_{n*}^{\zeta}\right)^{+} H_1 \xi_m^{\zeta'} \quad (20)$$

at $\boldsymbol{r} = (x, y)$, where $\phi$'s are bulk states for given ($k_x$, E), and $\phi_{n*}^{\zeta}$ denotes the state with $\mathbf{k}_{n^*} = \left( k_x, \left[ (k_y)_n^\zeta \right]^* \right)$. Moreover, for $\phi$'s of the same zone ($\zeta$), $\langle \phi_{n^*}^\zeta | J_y(\mathbf{r}) | \phi_m^\zeta \rangle \equiv J_{n^*,m}^\zeta(\mathbf{r})$ is diagonal, as shown in **Appendix II**.

To enforce the continuity condition, we apply the current operator on the local wave functions, one in zone $\zeta$ and the other in the neighboring zone ($\zeta'$), giving

$$J_y(\mathbf{R}=interface)\,\Psi^\zeta(\mathbf{R};k_x,E) = J_y(\mathbf{R}=interface)\,\Psi^{\zeta'}(\mathbf{R};k_x,E). \tag{21}$$

We project both sides of the equation onto $\phi_{n^*}^\zeta$, using the property of $J_{n^*,m}^\zeta$ being diagonal, and obtain the following transfer matrix equation

$$c_n^\zeta = \frac{1}{J_{n^*,m}^\zeta} \sum_m J_{n^*,m}^{\zeta,\zeta'}(\mathbf{R}) c_m^{\zeta'} \tag{22}$$

which expresses the linear combination coefficient in zone $\zeta$ in terms of those in the neighboring zone ($\zeta'$).

For the reflection/transmission problem, initial conditions are also required, e.g., $c_n^I = 1$ for the zone-I incident state, and $c_n^{II} = 0$ for the zone-II states traveling towards the interface. Together with these conditions, the last equation can be solved for $c_n^\zeta$'s and $c_n^{\zeta'}$'s. One can then evaluate the reflection and transmission coefficients by taking ratios between the coefficients, and investigate the valley contrast in electron transmission and reflection.

In our work, the contrast is measured by the quantities, $P_{trans}$ and $P_{reflec}$, the relative differences in the probabilities of finding the transmitted and reflected electrons in the two valleys, respectively. Specifically, $P_{trans}$ is defined in a way that depends on the incident electron energy relative to the barrier height as follows. In the case where the electron energy lies above the barrier (i.e., the first conduction band's minimum in zone II), at least four $(k_y)^{II}$'s are real, and $P_{trans}$ is given by

$$P_{trans}(\text{E}, k_x) = \frac{\sum_{\tau=+/-} \tau T^{(\tau)}}{\sum_{\tau=+/-} T^{(\tau)}}, \text{ where } T^{(\tau)} \equiv \left| \frac{j^{(\tau)}_{y,trans}}{j^{(\tau)}_{y,inc}} \right|,$$ *i.e.*, the ratio between the currents of the incident state and the $\tau$ valley component $\phi_n^{II,\tau}(\boldsymbol{r};,k_x,E)$ of the transmitted wave function $\Psi^{II}(\boldsymbol{r};k_x,E)$. For electron energy below the barrier, all $\left(k_y\right)^{II}$'s are complex with finite imaginary parts, and we define $P_{trans}$ in a similar way, but with $T^{(\tau)}$ re-defined as $T^{(\tau)} \equiv \sum_n \left| c_n^{II,\tau} \right|^2$, where '$n$' runs over the evanescent states from both the first and second conduction bands. In this case, although there are no propagating states below the barrier, the measure $P_{trans}$ as defined here is useful and relevant for the study of valley-dependent electron tunneling through barriers of finite width. For the reflection, we define and discuss the valley polarization $P_{reflec}$ as

$$P_{reflec} = \frac{\sum_{\tau=+/-} \tau R^{(\tau)}}{\sum_{\tau=+/-} R^{(\tau)}},$$ independent of the electron energy, where $R^{(\tau)} \equiv \left| \frac{j^{(\tau)}_{y,reflec}}{j^{(\tau)}_{y,inc}} \right|$, i.e. the ratio between the currents of the incident state and the reflected $\tau$ valley component $\phi_n^{I,\tau}(\boldsymbol{r};,k_x,E)$ of the wave function $\Psi^I(\boldsymbol{r};k_x,E)$. In calculating $P_{trans}$ and $P_{reflec}$, we focus on valley-conserving ($\tau \to \tau$) scattering events, and compare the results for the two valleys to evaluate the contrast. This comparison ignores the valley-flipping ($\tau \to -\tau$) events because they are very rare (See **Figure 3** below).

The formalism presented in this section allows us to study numerically both valley-conserving and valley-flipping electron transport in the structure, as discussed next in **Section III**.

### III.      Numerical Results and Discussions

We turn to the discussion of numerical results obtained with the formalism of **Subsection II-C**. Unless stated otherwise, we consider only **a)** incident electrons from the first conduction band in zone I (y < 0), and **b)** BLG based structures with zone II (y > 0) being the barrier, meaning the first conduction band's minimum in zone II is higher than that of zone I (y < 0).

First, we discuss the inter-valley coupling caused by the finite potential ($\delta V$) or gap discontinuity ($\delta \Delta$). **Figure 3** shows the interface-induced valley-flipping transmission, specifically, the ratio $\left|j_{rans}^{K'->K}\right|/\left|j_{rans}^{K'->K'}\right|$ as a function of electron energy for $\Delta^I = 10$ meV, $V^I = 0$, $\Delta^{II} = 20$ meV and $V^{II} = 20$ meV. It is found that the interface scattering gives rise to a limited amount of valley-flipping transmission far smaller than the valley-conserving one, typically by a factor of $10^{-5}$, consistent with the previous study of intervalley scattering in zigzag graphene nanoribbons[18]. This feature indicates that the potential fluctuation induced valley decoherence would be generally insignificant, making the valley-based electronics viable. In addition, the relative intervalley transfer is dependent on the electron energy in a monotonous fashion, but relatively insensitive to $k_x$. Specifically, as the electron energy increases, the potential or gap discontinuity eventually becomes ineffective in causing the intervalley transfer.

We then discuss the valley-conserving transport with various configurations of the interfaces. In **Figure 4**, we present the valley contrast for the transmission, $P_{trans}$, in the type-I structure specified by $\Delta^I = 10\ meV$, $V^I = 0$, $\Delta^{II} = 40\ meV$ and $V^{II} = 0$. The effect of the hopping $\gamma_3$ on valley contrast is investigated under different incident conditions, namely, (a) $k_x = 0$, (b) $k_x = 0.0035|K|$, and (c) $k_x = 0.007|K|$. For each condition, we plot two curves, one with the presence of $\gamma_3$ and the other with $\gamma_3 = 0$. It shows that the presence of $\gamma_3$ generally has sizable effects on the magnitude of the valley contrast, confirming the statement in **Subsection II-B** that warping lifts the valley degeneracy and alters the degree of contrast. Note, for the incident electron energy below the barrier height, it shows that the two results overlap in a major way. According to the discussion in **Subsection II-B**, we interpret the valley contrast shown here as largely caused by the existence of vertical interlayer coupling $\gamma_1$. In general, we see that the magnitude of the contrast depends on both $k_x$ and the electron energy. As we lower the energy and hence move into the tunneling regime, the magnitude increases. Moreover, it also increases with increasing $k_x$. In particular, for $k_x = 0.007|K|$, it is found that the polarization can reach a significant value of about 80% for electron energy around 10 meV. The sizable contrast demonstrated here for the tunneling states could have an important bearing on the development of valley filters, as it implies the likely existence of a promising valley filter in the tunneling regime.

In **Figure 5**, we present the valley contrast for the reflection, $P_{reflec}$, in the same type-I structure considered in **Figure 4**, for the incident electron energy higher than the barrier height. Note that the information provided by $P_{reflec}$ in **Figure 5** and that by $P_{trans}$ in **Figure 4** for the incident electron energy higher than the barrier height are complementary to each other because of the probability sum rule for the transmission and reflection coefficients, as reflected in the fact that $P_{trans}$ and $P_{reflec}$ are opposite in sign. However, they show interesting different dependences on the electron energy. While $P_{reflec}$ increases and saturates with increasing electron energy, $P_{trans}$ basically concentrates around the edge of the barrier and decreases with the energy. From the application perspective, the difference in the energy dependence as well as in the magnitude of the transmission / reflection coefficient (about unity for the transmission and small for the reflection, for over-barrier incident electrons) provides a flexibility in the design for valley polarized electron sources.

Last, we examine the relation between the valley contrast and the interface type, in **Figures 4**, **6**, and **7**, for type- I, II and III interfaces, respectively. The conduction band offsets in type- I and II are both taken to be 30 meV. We see that **Figures 4** and **6** show sizably different valley contrasts, indicating that the valley contrast magnitude of conduction band electrons is dependent on the valence band profile, with the type-I structure being more favorable than the type-II structure. Apart from the BLG-based structure, **Figure 6** also investigates a MLG-based structure, and shows a small magnitude, e.g., $10^{-7}$, of warping-based valley contrast for general $k_x$. For type-III structures, **Figure 7** shows that the contrast can be quite significant and even greater than that in the type-I structure. In connection with the recent studies of valley polarization in graphene-based quantum structures, we note that BLG-based structures with inverted gap alignment have been theoretically proposed by Morpurgo et al.[9], and experimentally studied by Wang et al.[19] using grain boundaries as interfaces between AB and BA stacked zones. These studies provide promising suggestions for applications in valleytronics. However, they focused on the quasi-one-dimensional (Q1D) channel formed along the interface and investigated the K / K′ chiral states confined and propagating in the channel. In contrast, the present work looks into the valley dependence of bulk electron scattering by the interface, for both normal and oblique incidences and, in the case of an inverted gap alignment, our analysis finds large valley polarization, similar in magnitude to that reported in the two aforementioned studies.

## IV. Conclusion

In summary, we have presented a multi-band theory of electronic states for valley dependent electron transport in both MLG- and BLG- based lateral quantum structures. In particular, a symmetry-based analysis has been performed, within the linearized model, for structures with one interface. In the MLG case, it shows that the valley contrast in the reflection or transmission exists only when going beyond the linearized approximation, e.g., by including the band warping in the model. This nonlinear origin of valley contrast severely limits the magnitude of the effect except for high energy electrons. In comparison, in the BLG case the analysis suggests, even in the linearized approximation, the existence of valley contrast due to both the warping and vertical interlayer coupling. This therefore results in the occurrence of a sizable contrast at low electron energy, making BLG a favorable valleytronic system.

In addition, we have developed a full TB theoretical formalism with the usage of a cell-averaged current operator to construct the global wave function from local bulk states. This formalism allows us to include the interface-induced intervalley scattering in the study, and is well suited for the numerical investigation of valley contrast. The calculations with three types of interfaces support the conclusion of the linearized model-based analysis and, moreover, identify the vertical interlayer coupling as the main cause for valley contrast in the electron tunneling transport in BLG-based structures. The result also demonstrates the interesting dependence of valley polarization on the parallel momentum $k_x$.

In conclusion, our theoretical study with the TB model shows the existence of sizable valley contrast in the graphene-based structures with in-plane gap or potential variations - even for those simple configurations with only one interface. In view of the feasibility to create versatile gap or potential variations, for example, in BLG with state-of-the-art gating technology, the study reported in this work has an intriguing implication for the implementation of valleytronics with graphene-based lateral quantum structures.

*Acknowledgment*

We would like to thank the financial support of MoST, ROC through the Contract No. 103-2119-M-007-007-MY3. M.Y.C. acknowledges the support from the Academia Sinica Thematic Project and NSF Grant No. 1542747.

# Appendix I
## The cell-averaged current operator

Here, we derive the current matrix element $\langle \phi_1 | J_y(\vec{R}) | \phi_2 \rangle$ in the MLG case. A similar derivation in the BLG case leads to the same expression.

**The Result:**

$$\langle \phi_1 | J_y(\vec{R_0}) | \phi_2 \rangle = \frac{i}{\hbar} \frac{\sqrt{3}}{8} \frac{a}{N} \exp\left\{ i \left[ (k_y)_2 - (k_y)_1^* \right] y_0 \right\} \left( \lambda_1^* - \lambda_1^{*-1} + \lambda_2 - \lambda_2^{-1} \right) \xi_1^+ H_1 \xi_2 \quad \text{(A1)}$$

$J_y(\vec{R_0})$: the y-component of the current density operator averaged over the unit cell at $\vec{R_0}$.

$|\phi_m\rangle$: the bulk eigenstate with wave vector $\vec{k_m}$. Within the present TB model of graphene, we write the bulk eigenstate with wave vector $k_m$ in the form:

$$|\phi_m\rangle = \sum_{\vec{l}\, \in\, \text{all atomic sites}} b_{l,m} |\vec{l}\rangle, \quad \text{(A2)}$$

where $|\vec{l}\rangle$ is the atomic orbital at site $\vec{l}$.

$$\xi_m = \left( \langle \Phi_A(\vec{k}_m) | \phi_m \rangle, \; \langle \Phi_B(\vec{k}_m) | \phi_m \rangle \right)^T, \; (k_y)_m = \left( \vec{k_m} \right)_y, \; \lambda_m = e^{i\sqrt{3}(k_y)_m a/2}. \quad \text{(A3)}$$

$\Phi_A(\vec{k}_m)$ and $\Phi_B(\vec{k}_m)$ are the Bloch basis functions, composed of A site and B site orbitals, respectively. (See below.)

$H_1$ refers to the matrix in the $k_y$-dependent part of the Hamiltonian, as defined by the expression

$$H\left(\vec{k_m}\right) = H_0(k_x) + (\lambda_m + \lambda_m^{-1}) H_1(k_x). \quad \text{(A4)}$$

**The Derivation:**

In the discrete model such as the present one, there are two kinds of currents - the link current and the point current[20]. The link current flows on the linkage line between atoms; while the point current is what is observed at an atomic site. In this derivation, we focus on the point current, since our theoretical formalism is concerned with matching currents at any point of the interface when joining local wave functions. We list below a few fundamental facts required for the evaluation of point current:

(1) The $\mu$ component ($\mu$ = x or y) of point current density operator at *atomic site* $\vec{l}$ is given by

$$j_\mu(\vec{l}) = \frac{1}{2}\{n_l \hat{v}_\mu + \hat{v}_\mu n_l\} \tag{A5}$$

(2) The point density operator at atomic site $\vec{l}$ is given by

$$n_l = |\vec{l}\rangle\langle\vec{l}|. \tag{A6}$$

(3) The velocity operator is given by $v_\mu = \frac{1}{i\hbar}[\hat{r}_\mu, H]$. \hfill (A7)

(4) The matrix element of the posistion operator is taken to be stricktly diagonal:

$$\langle\vec{l'}|\hat{r}_\mu|\vec{l}\rangle = (\vec{l})_\mu \delta_{\vec{l},\vec{l'}} \tag{A8}$$

The evaluation of the matrix element of the point current $j_y(\vec{l})$ between bulk states $|\phi_1\rangle$ and $|\phi_2\rangle$ yields

$\langle\phi_1|j_y(\vec{l})|\phi_2\rangle = \sum_{l'}(j_y)_{1,2}(\vec{l'},\vec{l})$, where

$$(j_y)_{1,2}(\vec{l'},\vec{l}) = \frac{1}{2i\hbar}(\vec{l'}-\vec{l})_y\{b_{l,2}(b_{l',1})^*\langle l'|H|l\rangle - b_{l',2}(b_{l,1})^*\langle l|H|l'\rangle\}. \tag{A9}$$

In passing, we note that the partial current $(j_y)_{1,2}(\vec{l'},\vec{l})$ equals one half of the link current from atomic site $\vec{l}$ to $\vec{l'}$ [20]. We apply the last expression to graphene. With only the nearest neighbor (n.n.) hopping considered in the present model, we obtain, for example, the A site point current at $\vec{R}_0$ as

$\langle\phi_1|j_y(\vec{R}_0)|\phi_2\rangle = \sum_{m=1,2,3}(j_y)_{1,2}(\vec{R}_0+\vec{\delta}_m,\vec{R}_0)$, where

$$(j_y)_{1,2}(\vec{R}_0+\vec{\delta}_m,\vec{R}_0) = \frac{1}{2i\hbar}(\vec{\delta}_m)_y\left[b_{\vec{R}_0,2}\left(b_{\vec{R}_0+\vec{\delta}_m,1}\right)^*\langle\vec{R}_0+\vec{\delta}_m|H|\vec{R}_0\rangle \right.$$
$$\left. -b_{\vec{R}_0+\vec{\delta}_m,2}\left(b_{\vec{R}_0,1}\right)^*\langle\vec{R}_0|H|\vec{R}_0+\vec{\delta}_m\rangle\right] \tag{A10}$$

In the above, $\delta_m$ (m = 1, 2, 3) specifies the n.n. position of each atom, and $\langle \vec{R_0}|H|\vec{R_0}+\vec{\delta_m}\rangle = \langle \vec{R_0}+\vec{\delta_m}|H|\vec{R_0}\rangle^* = -t^* = -t$ (the hopping).

For further evaluation, we specify explicitly the atomic sites, the bulk states $|\phi_1\rangle$ and $|\phi_2\rangle$ in graphene, and the amplitudes $b_l$'s, as follows. Let $a$ = C-C distance. The positions of the two basis atoms in a unit cell at $\vec{R_0}$ are A: $\vec{R_0}$ = ($x_0$, $y_0$) and B: $\vec{R_0}$ + $a$ (1, 0). The nearest neighbor (n.n.) B sites of A are located at $\vec{R_0}+\vec{\delta_1}$, $\vec{R_0}+\vec{\delta_2}$, and $\vec{R_0}+\vec{\delta_3}$, where $\delta_1 = a$ (1, 0), $\delta_2 = a$ (-1/2, $\sqrt{3}$/2), and $\delta_3 = a$ (-1/2, -$\sqrt{3}$/2). The n.n. A sites of B are given by $\vec{R_0}, \vec{R_0}+\vec{\delta_1}-\vec{\delta_3}$ and $\vec{R_0}+\vec{\delta_1}-\vec{\delta_2}$. The Bloch wave function is given by $|\phi_m\rangle = c_{A,m}\Phi_A(\vec{k}_m) + c_{B,m}\Phi_B(\vec{k}_m)$, in terms of the Bloch basis functions $\Phi_A(\vec{k}_m)$ and $\Phi_B(\vec{k}_m)$. $\Phi_A(\vec{k}_m)$ and $\Phi_B(\vec{k}_m)$ are given by

$$\Phi_A(\vec{k}_m) = \frac{1}{\sqrt{N}} \sum_{\vec{R}\in \text{lattice vectors}} e^{i\vec{k}_m\cdot\vec{R}}\varphi_a(\vec{r}-\vec{R}),  \quad (\text{A11a})$$

$$\Phi_B(\vec{k}_m) = \frac{1}{\sqrt{N}} \sum_{\vec{R}\in \text{lattice vectors}} e^{i\vec{k}_m\cdot\vec{R}}\varphi_a(\vec{r}-\vec{R}-\vec{\delta_1}).  \quad (\text{A11b})$$

N is the number of unit cells, and $\varphi_a(\vec{r}-\vec{R}) = \langle \vec{r}|\vec{l}=\vec{R}\rangle$ is the $2p_z$ orbital at $\vec{R}$. $b_{l,m} = \frac{1}{\sqrt{N}} c_{A,m} e^{i\vec{k}\cdot\vec{l}}$ for $\vec{l} \in$ A sites, and $b_{l,m} = \frac{1}{\sqrt{N}} c_{B,m} e^{i\vec{k}\cdot(\vec{l}-\vec{\delta_1})}$ for $\vec{l} \in$ B sites.

With $b_{l,m}$'s above being substituted into $(j_y)_{1,2}(\vec{R_0},\vec{R_0}+\vec{\delta_m})$, we obtain

$$(j_y)_{1,2}(\vec{R_0}+\vec{\delta_1},\vec{R_0}) = 0, \text{ since } (\vec{\delta_1})_y = 0. \quad (\text{A12a})$$

$$(j_y)_{1,2}(\vec{R_0}+\vec{\delta_2},\vec{R_0}) = \frac{1}{2i\hbar}\left(\frac{\sqrt{3}}{2}\right)[-at]\frac{1}{N}\exp(i\vec{k}_2\cdot\vec{R_0}-i\vec{k}_1^*\cdot\vec{R_0})$$

$$\times\left\{c_{A,2}e^{-i\vec{k}_1^*\cdot(-\vec{\delta_1}+\vec{\delta_2})}c_{B,1}^* - e^{i\vec{k}_2\cdot(-\vec{\delta_1}+\vec{\delta_2})}c_{B,2}c_{A,1}^*\right\}, \quad (\text{A12b})$$

$$(j_y)_{1,2}(\overrightarrow{R_0}+\overrightarrow{\delta_3},\overrightarrow{R_0}) = \frac{-1}{2i\hbar}\left(\frac{\sqrt{3}}{2}\right)[-at]\frac{1}{N}\exp(i\vec{k}_2\cdot\vec{R}_0 - i\vec{k}_1^*\cdot\vec{R}_0)$$

$$\times\left\{c_{A,2}e^{-i\vec{k}_1^*\cdot(-\vec{\delta}_1+\vec{\delta}_3)}c_{B,1}^* - e^{i\vec{k}_2\cdot(-\vec{\delta}_1+\vec{\delta}_3)}c_{B,2}c_{A,1}^*\right\}. \quad (A12c)$$

In our application, we require $(k_1)_x = (k_2)_x = k_x$ (a real number) for the interface-scattered state, which allows for further simplification for $(j_y)_{1,2}(\overrightarrow{R_0}+\overrightarrow{\delta_2},\overrightarrow{R_0})$ and $(j_y)_{1,2}(\overrightarrow{R_0}+\overrightarrow{\delta_3},\overrightarrow{R_0})$:

$$(j_y)_{1,2}(\overrightarrow{R_0}+\overrightarrow{\delta_2},\overrightarrow{R_0}) = \frac{1}{2i\hbar}\left(\frac{\sqrt{3}}{2}\right)[-at]\frac{1}{N}\exp\left\{i\left[(k_y)_2 - (k_y)_1^*\right]y_0\right\}$$

$$\times\left\{e^{i3k_xa/2}\lambda_1^* c_{A,2}c_{B,1}^* - e^{-i3k_xa/2}\lambda_2 c_{B,2}c_{A,1}^*\right\}, \quad (A13a)$$

$$(j_y)_{1,2}(\overrightarrow{R_0}+\overrightarrow{\delta_3},\overrightarrow{R_0}) = \frac{-1}{2i\hbar}\left(\frac{\sqrt{3}}{2}\right)\frac{1}{N}[-at]\exp\left\{i\left[(k_y)_2 - (k_y)_1^*\right]y_0\right\}$$

$$\times\left\{e^{i3ak_x/2}\lambda_1^{*-1} c_{A,2}c_{B,1}^* - e^{-i3ak_x/2}\lambda_2^{-1} c_{B,2}c_{A,1}^*\right\}. \quad (A13b)$$

Similarly, we obtain the point current at B site located at $\overrightarrow{R_0}+\overrightarrow{\delta_1}$:

$$\langle\phi_1|j_y(\overrightarrow{R_0}+\overrightarrow{\delta_1})|\phi_2\rangle = \sum_{m=1,2,3}(j_y)_{1,2}(\overrightarrow{R_0}+\overrightarrow{\delta_1}-\overrightarrow{\delta_m},\overrightarrow{R_0}+\overrightarrow{\delta_1}), \text{ where}$$

$$j_y^{(1,2)}(\overrightarrow{R_0},\overrightarrow{R_0}+\overrightarrow{\delta_1}) = 0, \quad (A14a)$$

$$(j_y)_{1,2}(\overrightarrow{R_0}+\overrightarrow{\delta_1}-\overrightarrow{\delta_2},\overrightarrow{R_0}+\overrightarrow{\delta_1}) = \frac{-1}{2i\hbar}\left(\frac{\sqrt{3}}{2}\right)\frac{1}{N}[-at]\exp\left\{i\left[(k_y)_2 - (k_y)_1^*\right]y_0\right\}$$

$$\times\left\{e^{-i3k_xa/2}\lambda_1^{*-1} c_{B,2}c_{A,1}^* - e^{i3k_xa/2}\lambda_2^{-1} c_{A,2}c_{B,1}^*\right\}, \quad (A14b)$$

$$(j_y)_{1,2}(\overrightarrow{R_0}+\overrightarrow{\delta_1}-\overrightarrow{\delta_3},\overrightarrow{R_0}+\overrightarrow{\delta_1}) = \frac{1}{2i\hbar}\left(\frac{\sqrt{3}}{2}\right)\frac{1}{N}[-at]\exp\left\{i\left[(k_y)_2 - (k_y)_1^*\right]y_0\right\}$$

$$\times\left\{e^{-i3k_xa/2}\lambda_1^* c_{B,2}c_{A,1}^* - e^{i3k_xa/2}\lambda_2 c_{A,2}c_{B,1}^*\right\}. \quad (A14c)$$

The cell-averaged $\langle \phi_1 | J_y(\vec{R}_0) | \phi_2 \rangle$ is evaluated by taking the average of the current densities at A and B sites:

$$\langle \phi_1 | J_y(\vec{R}_0) | \phi_2 \rangle = \frac{1}{2}\left[\langle \phi_1 | j_y(\vec{R}_0) | \phi_2 \rangle + \langle \phi_1 | j_y(\vec{R}_0 + \vec{\delta}_1) | \phi_2 \rangle\right]$$

$$= \frac{i}{4\hbar}\frac{\sqrt{3}}{2}\frac{1}{N}[-at]\exp\left\{i\left[(k_y)_2 - (k_y)_1^*\right]y_0\right\}\left[-(\lambda_1^* - \lambda_1^{*-1}) + (\lambda_2 - \lambda_2^{-1})\right]$$

$$\times \left[e^{i3k_xa/2}c_{A,2}c_{B,1}^* + e^{-i3k_xa/2}c_{B,2}c_{A,1}^*\right]. \qquad (A15)$$

In order to obtain the final form of $\langle \varphi_1 | J_y(\vec{R}_0) | \varphi_2 \rangle$, we list below the MLG Hamiltonian matrix:

$$H(k_y;k_x) = H_0(k_x) + \eta H_1(k_x), \ \eta = \lambda + \lambda^{-1}, \lambda \equiv \exp(i\sqrt{3}k_ya/2), \qquad (A16a)$$

$$H = \begin{pmatrix} -\Delta & -t\left[1+\eta e^{-i3k_xa/2}\right] \\ -t\left[1+\eta e^{i3k_xa/2}\right] & \Delta \end{pmatrix}, H_0 = \begin{pmatrix} -\Delta & -t \\ -t & \Delta \end{pmatrix}, H_1 = \begin{pmatrix} 0 & -te^{-i3k_xa/2} \\ -te^{i3k_xa/2} & 0 \end{pmatrix},$$

(A16b)

with the eigenstate of $H$ being given by $\xi_m = \begin{pmatrix} c_{A,m}, & c_{B,m} \end{pmatrix}^T$. It follows that

$$\xi_1^+ H_1 \xi_2 = \begin{pmatrix} c_{A,1}^* & c_{B,1}^* \end{pmatrix}\begin{pmatrix} 0 & -te^{-i3k_xa/2} \\ -te^{i3k_xa/2} & 0 \end{pmatrix}\begin{pmatrix} c_{A,2} \\ c_{B,2} \end{pmatrix} = -t\left[e^{-i3k_xa/2}c_{B,2}c_{A,1}^* + e^{i3k_xa/2}c_{A,2}c_{B,1}^*\right]$$

(A17)

Thus, in terms of the notation of $H_1$, we can alternatively write

$$\langle \phi_1 | J_y(\vec{R}_0) | \phi_2 \rangle = \frac{i}{\hbar}\frac{\sqrt{3}}{8}\frac{a}{N}\exp\left\{i\left[(k_y)_2 - (k_y)_1^*\right]y_0\right\}\left[-(\lambda_1^* - \lambda_1^{*-1}) + (\lambda_2 - \lambda_2^{-1})\right]\xi_1^+ H_1 \xi_2.$$

(A18)

# Appendix II

## Orthogonality of bulk eigenstates with respect to the current operator

Let

$$J_{1*,2} = <\phi_{1*} | J_y(\boldsymbol{R}_0) | \phi_2>$$
$$= \frac{i}{\hbar} \frac{\sqrt{3}}{8} \frac{a}{N} \exp\left\{i\left[(k_y)_2 - (k_y)_1\right] y_0\right\} \left[(\lambda_1 - \lambda_1^{-1}) + (\lambda_2 - \lambda_2^{-1})\right] \xi_{1*}^{+} H_1 \xi_2 \quad \text{(B1)}$$

$|\phi_{1*}\rangle$ and $|\phi_2\rangle$ are the bulk eigenstates characterized by $\left(k_x, (k_y)_1^*, E\right)$ and $\left(k_x, (k_y)_2, E\right)$ respectively. We show $J_{1*,2} = 0$ for $(k_y)_1 \neq (k_y)_2$.

**Proof:**

$$0 = \int d^2 r \left[\phi_{1*}^{+} H \phi_2 - (H\phi_{1*})^{+} \phi_2\right] \quad \text{(B2a)}$$

$$= \frac{1}{N} \sum_{\boldsymbol{R}} \exp\left\{i\left[(k_y)_2 - (k_y)_1\right](\boldsymbol{R})_y\right\}$$
$$\times \xi_{1*}^{+} \left\{\left[H_0(k_x) + (\lambda_2 + \lambda_2^{-1}) H_1(k_x)\right] - \left[H_0(k_x) + (\lambda_1 + \lambda_1^{-1}) H_1(k_x)\right]\right\} \xi_2 \quad \text{(B2b)}$$

$$= \frac{1}{N} \sum_{\boldsymbol{R}} \exp\left[i(k_{y,2} - k_{y,1})(\boldsymbol{R})_y\right] \left[\xi_{1*}^{+} (\lambda_2 + \lambda_2^{-1} - \lambda_1 - \lambda_1^{-1}) H_1(k_x) \xi_2\right] \quad \text{(B2c)}$$

Therefore,

$$0 = \xi_{1*}^{+} H_1(k_x) \xi_2 \text{ for } \lambda_1 \neq \lambda_2 \left[\text{i.e. } (k_y)_1 \neq (k_y)_2\right] \wedge \lambda_1 \neq \lambda_2^{-1} \left[\text{i.e. } (k_y)_1 \neq -(k_y)_2\right] \quad \text{(B3)}$$

Although for $\lambda_1 = \lambda_2^{-1}$ (i.e. $(k_y)_1 = -(k_y)_2$), $\xi_{1*}^{+} H_1(k_x) \xi_2$ is not necessarily zero, the orthogonality of $<\phi_{1*} | j_y(\boldsymbol{R}) | \phi_2>$ still holds since the term $\left[(\lambda_1 - \lambda_1^{-1}) + (\lambda_2 - \lambda_2^{-1})\right]$ inside the current matrix element vanishes in this case.

# Figure Captions

**Figure 1** (a) MLG and AB-stacked BLG showing lattice vectors and sub-lattices. Each unit cell consists of two carbon atoms A1 and B1 in the MLG case, or four carbon atoms A1 and B1 in the first layer and A2 and B2 in the second layer in the BLG case. (b) The corresponding Brillouin zone. (c) The bulk BLG energy contours of the first conduction band at two different energies $E_1$ and $E_2$, which illustrate the trigonal band warping around the Dirac point. Here, $k_x^{norm} = k_x / |K|$ and $k_y^{norm} = k_y / |K|$. The parameters are given by $t = 2.8$ eV, $\Delta = 10$ meV, $E_1 = 30$ meV and $E_2 = 50$ meV, where $E_1$ and $E_2$ correspond to the inner and outer energy contour lines, respectively.

**Figure 2** The complex band structure $E(k_y)$ of bulk BLG with $k_x = 0.0035|K|$ and $\Delta = 20$ meV. $k_y^{norm} = k_y / |K|$. (a) shows the existence of complex $k_y$'s in the energy range between the first and second conduction bands, (b) blows up the portion of the band structure close to the first conduction band edge, and (c) shows the calculation in the linearized model with $\gamma_3 = 0$.

**Figure 3** The interface-induced valley-flipping transmission in a type-II structure with parameters specified by $\Delta^I = 10$ meV, $V^I = 0$, $\Delta^{II} = 20$ meV and $V^{II} = 20$ meV. Specifically, we plot the ratio $\left| j_{rans}^{K' \to K} / j_{rans}^{K' \to K'} \right|$ as a function of electron energy. Three cases are investigated: normal incidence ($k_x = 0$) and oblique incidences ($k_x = 0.0035|K|$ and $0.007|K|$).

**Figure 4** The valley contrast for the transmission - $P_{trans}$ in the type-I structure specified by $\Delta^I = 10$ meV, $V^I = 0$, $\Delta^{II} = 40$ meV and $V^{II} = 0$. The effect of the hopping $\gamma_3$ on valley polarization is investigated under three different incident conditions, namely, (a) $k_x = 0$, (b) $k_x = 0.0035|K|$, and (c) $k_x = 0.007|K|$. For each condition, we plot two curves, one with the presence of $\gamma_3$ and the other with $\gamma_3 = 0$. The vertical dotted line indicates the conduction band edge in zone II, and varies with $k_x$.

**Figure 5** The valley contrast for the reflection - $P_{reflec}$ for the incident electron energy higher than the barrier height, in the type-I structure specified by $\Delta^{I} = 10$ meV, $V^{I} = 0$, $\Delta^{II} = 40$ meV and $V^{II} = 0$. The effect of the hopping $\gamma_3$ on valley polarization is investigated under two different incident conditions, namely, **(a)** $k_x = 0.0035|K|$ and **(b)** $k_x = 0.007|K|$. For each condition, we plot two curves, one with the presence of $\gamma_3$ and the other with $\gamma_3 = 0$.

**Figure 6** $P_{trans}$ as a function of incident electron energy and transverse wave vector $k_x$, for both MLG- and BLG- based type-II structures. **(a)** For the MLG-based structure, we take $\Delta^{I} = 20$ meV, $V^{I} = 0$, $\Delta^{II} = 20$ meV, $V^{II} = 30$ meV, and $k_x = 0$, 0.0035|K| and 0.007|K|. Both the conduction and valence band offsets are taken to be 30meV. **(b)** For the BLG-based structure, we take $\Delta^{I} = 10$ meV, $V^{I} = 0$, $\Delta^{II} = 20$ meV, $V^{II} = 20$ meV, and $k_x = 0.0035|K|$ and 0.007|K|. The conduction and valence band offsets are taken to be 30 meV and 10 meV, respectively.

**Figure 7** $P_{trans}$ as a function of incident electron energy and transverse wave vector $k_x$, with $k_x = 0.0035|K|$ and 0.007|K|, in an inverted structure with parameters specified by $\Delta^{I} = 25$ meV, $V^{I} = 0$, $\Delta^{II} = -25$ meV, $V^{II} = 0$. The conduction and valence band offsets are both taken to be 0 meV.

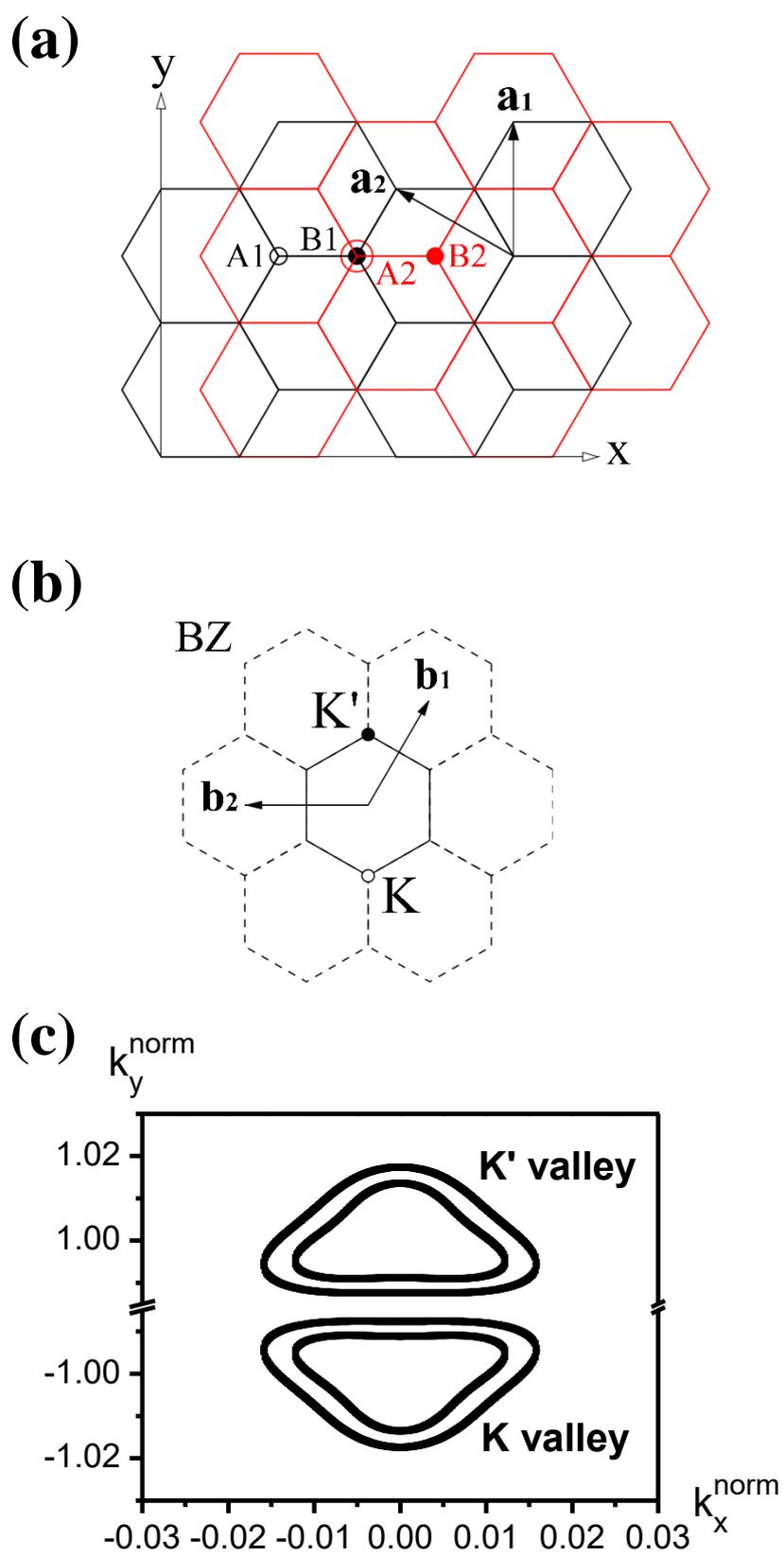

**Figure 1**

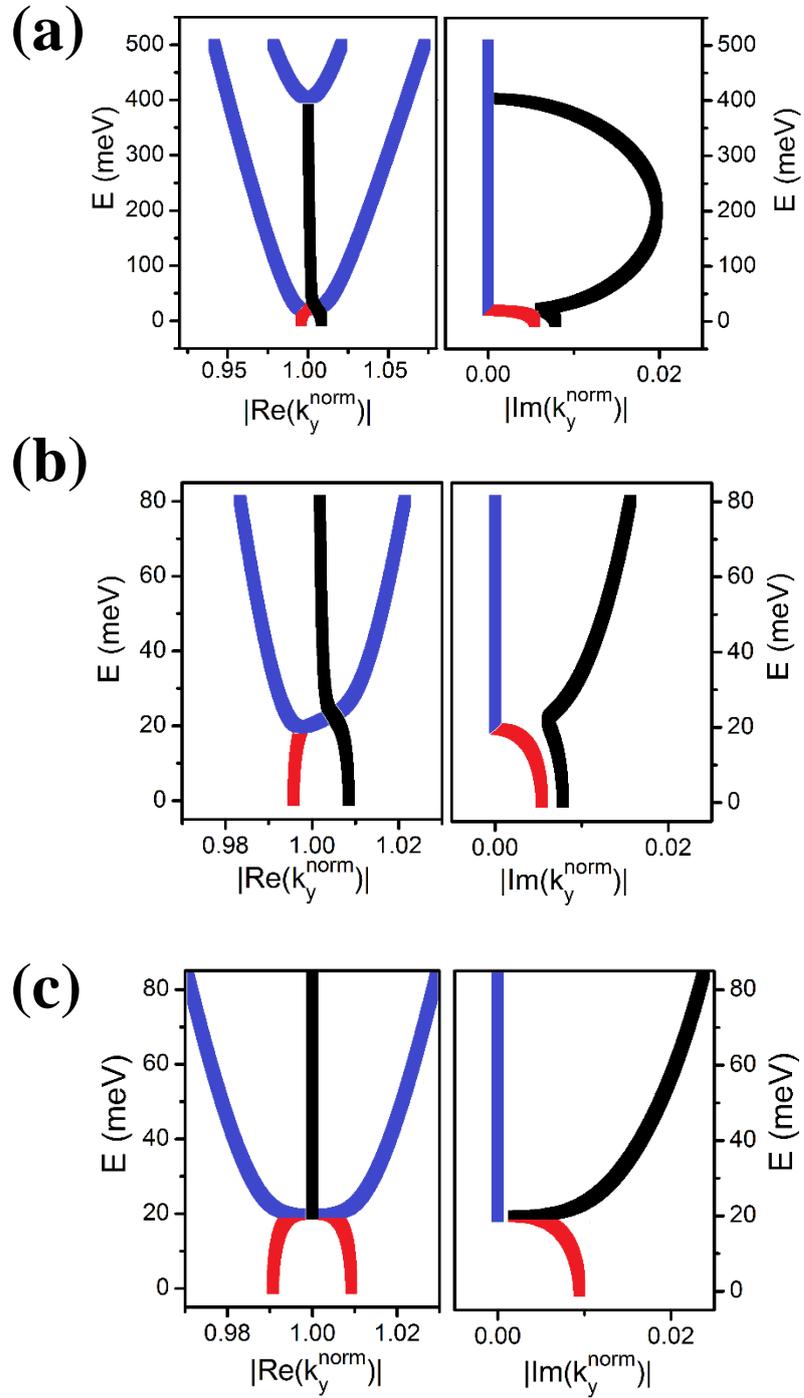

**Figure 2**

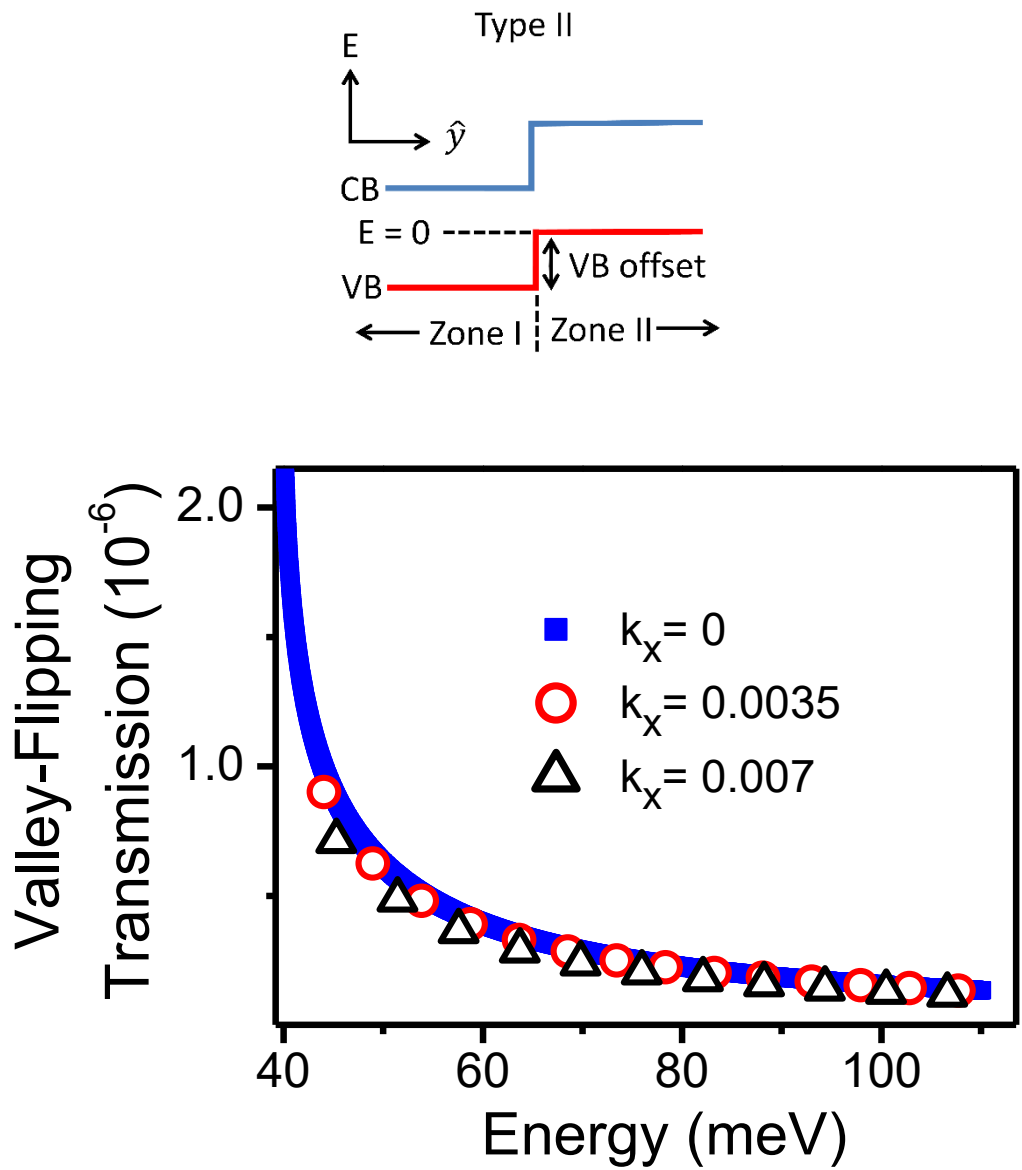

**Figure 3**

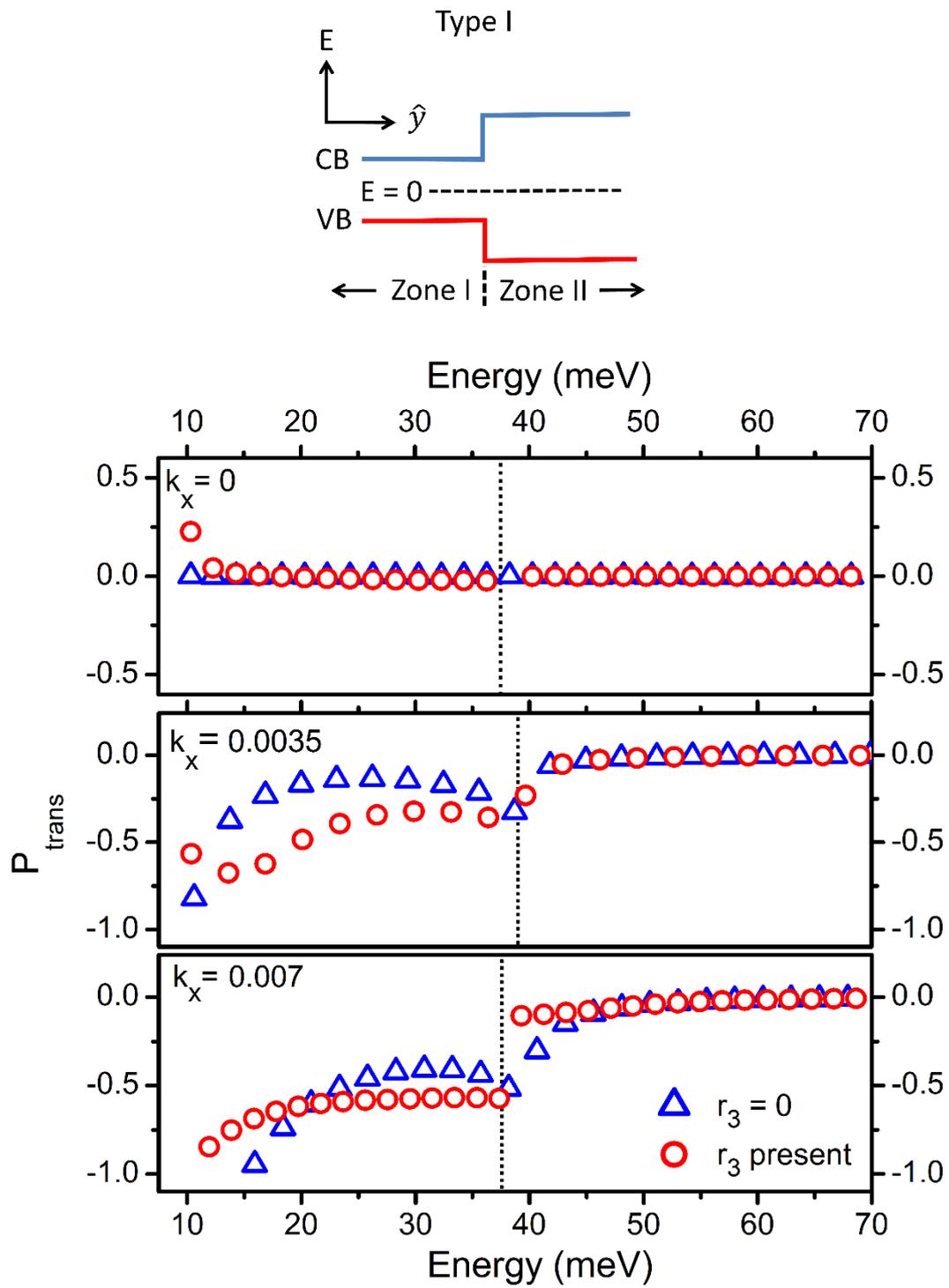

Figure 4

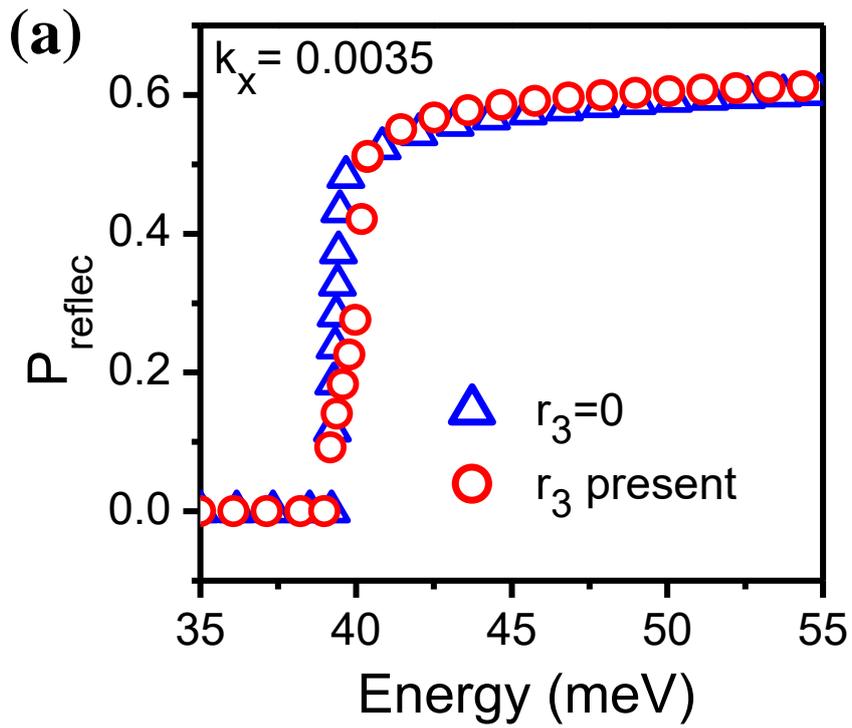

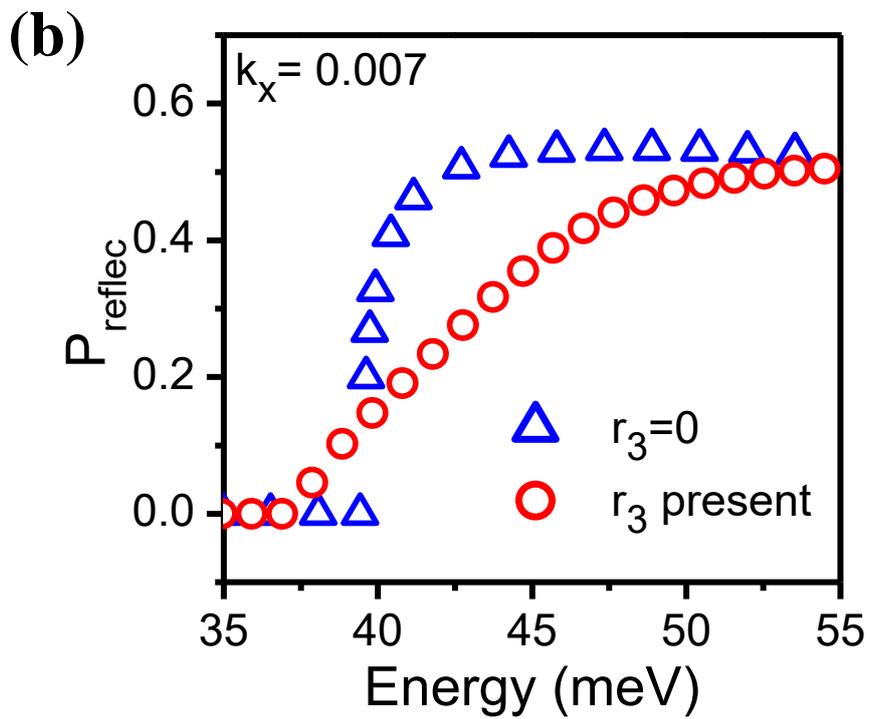

**Figure 5**

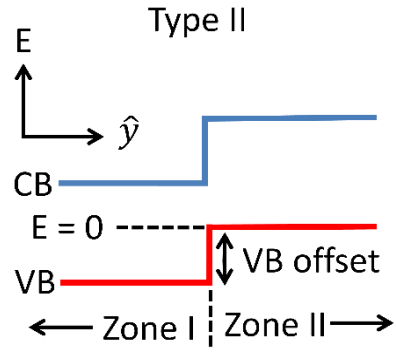

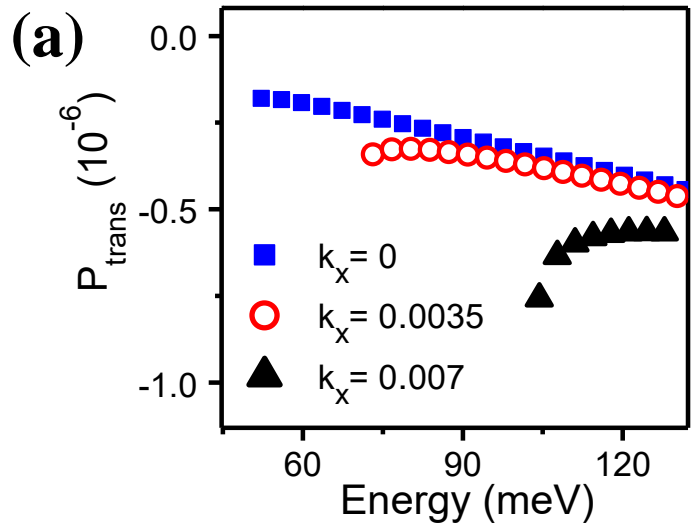

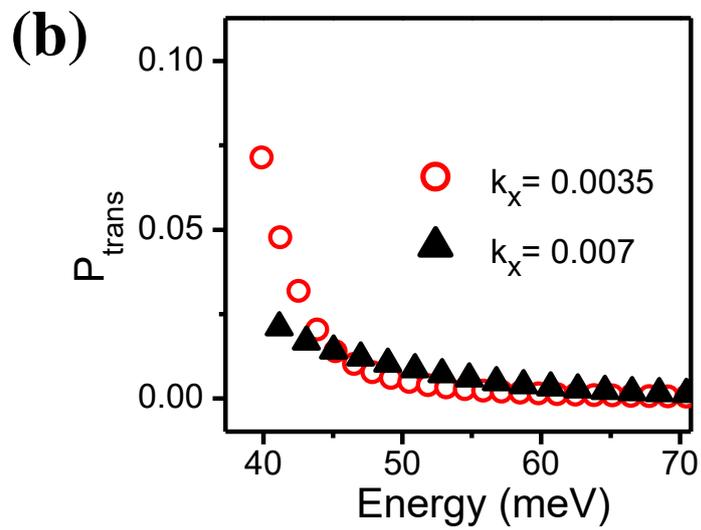

**Figure 6**

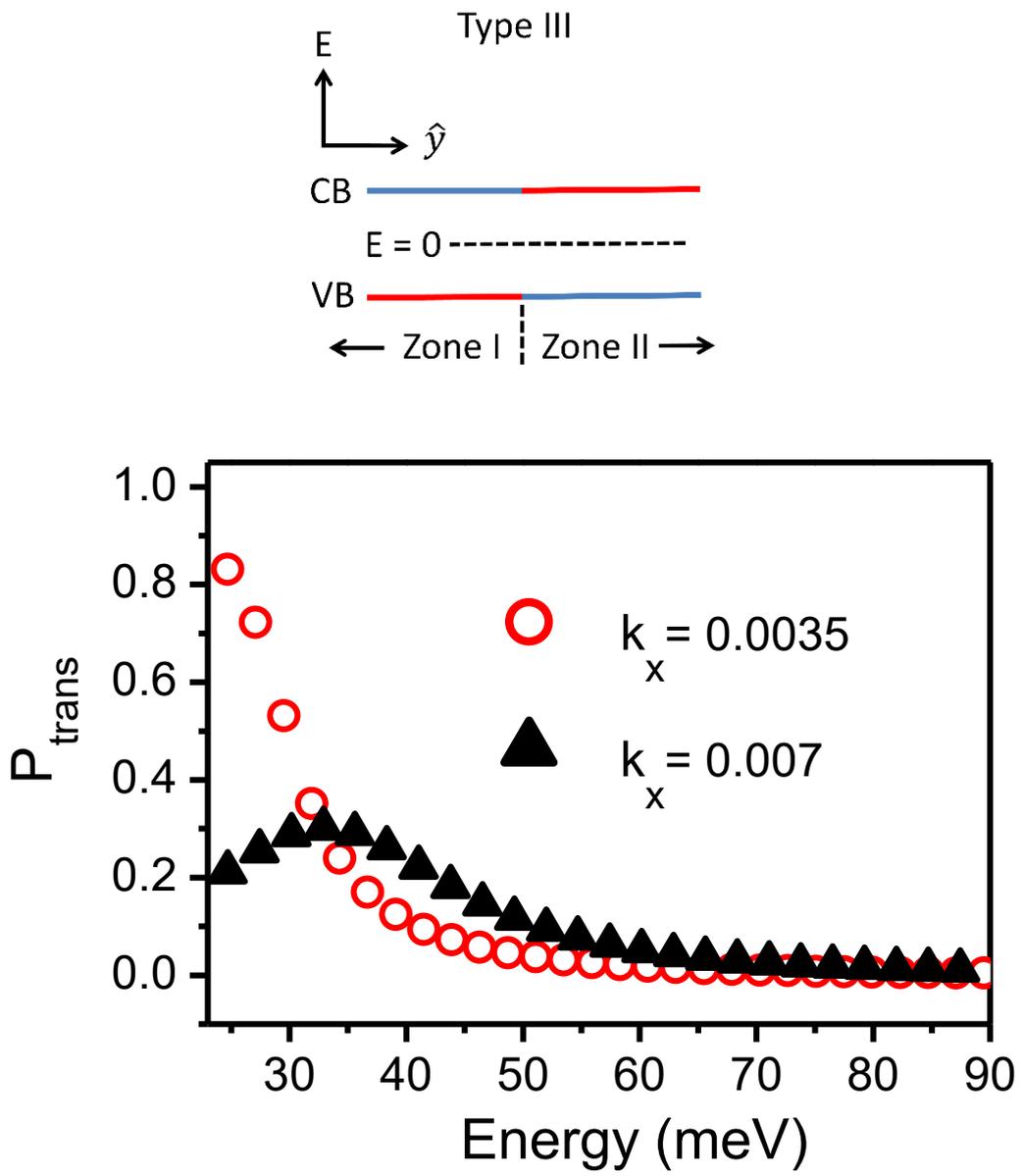

**Figure 7**